\documentclass[%
reprint,
superscriptaddress,
 amsmath,amssymb,
 aps,
prb,
floatfix,
]{revtex4-1}

\usepackage[utf8]{inputenc}
\usepackage{color,graphicx,soul}
\usepackage{dcolumn}
\usepackage{bm}
\usepackage{subfigure}

\begin{document}

\preprint{APS/123-QED}

\title{Optical Properties of Graphene Nanoflakes: Shape Matters}

\author{Candela Mansilla Wettstein}
\affiliation{Instituto de Investigaciones Fisicoquímicas de Córdoba, Consejo Nacional de Investigaciones Científicas y Técnicas (INFIQC - CONICET), Departamento de Matemática y Física, Facultad de Ciencias Químicas, Universidad Nacional de Córdoba, X5000HUA, Ciudad Universitaria, Córdoba, Argentina.}
\author{Franco P. Bonafé}
\affiliation{Instituto de Investigaciones Fisicoquímicas de Córdoba, Consejo Nacional de Investigaciones Científicas y Técnicas (INFIQC - CONICET), Departamento de Matemática y Física, Facultad de Ciencias Químicas, Universidad Nacional de Córdoba, X5000HUA, Ciudad Universitaria, Córdoba, Argentina.}
\author{M. Bel\'en Oviedo}
\affiliation{Department of Chemical \& Environmental Engineering and Materials Science and Engineering Program, University of California, Riverside, California 92521, United States}
\author{Cristián G. Sánchez}
\email{cgsanchez@fcq.unc.edu.ar}
\affiliation{Instituto de Investigaciones Fisicoquímicas de Córdoba, Consejo Nacional de Investigaciones Científicas y Técnicas (INFIQC - CONICET), Departamento de Matemática y Física, Facultad de Ciencias Químicas, Universidad Nacional de Córdoba, X5000HUA, Ciudad Universitaria, Córdoba, Argentina.}

\date{\today}

\begin{abstract}
In recent years there has been significant debate on whether the edge type of graphene nanoflakes (GNF) or graphene quantum dots (GQD) are relevant for their electronic structure, thermal stability and optical properties. Using computer simulations, we have proven that there is a fundamental difference in the absorption spectra between samples of the same shape, similar size but different edge type, namely, armchair or zigzag edges. These can be explained by the presence of electronic structures near the Fermi level which are localized on the edges. These features are also evident from the dependence of band gap on the GNF size, which shows three very distinct trends for different shapes and edge geometries.

\end{abstract}

\maketitle

\begin{figure}[ht]
\subfigure{\includegraphics[width=40mm]{TAC-espect}}
\subfigure{\includegraphics[width=40mm]{TZZ-espect-SP}}
\subfigure{\includegraphics[width=40mm]{HAC-espect}}
\subfigure{\includegraphics[width=40mm]{HZZ-espect-SP}}
\subfigure{\includegraphics[width=80mm]{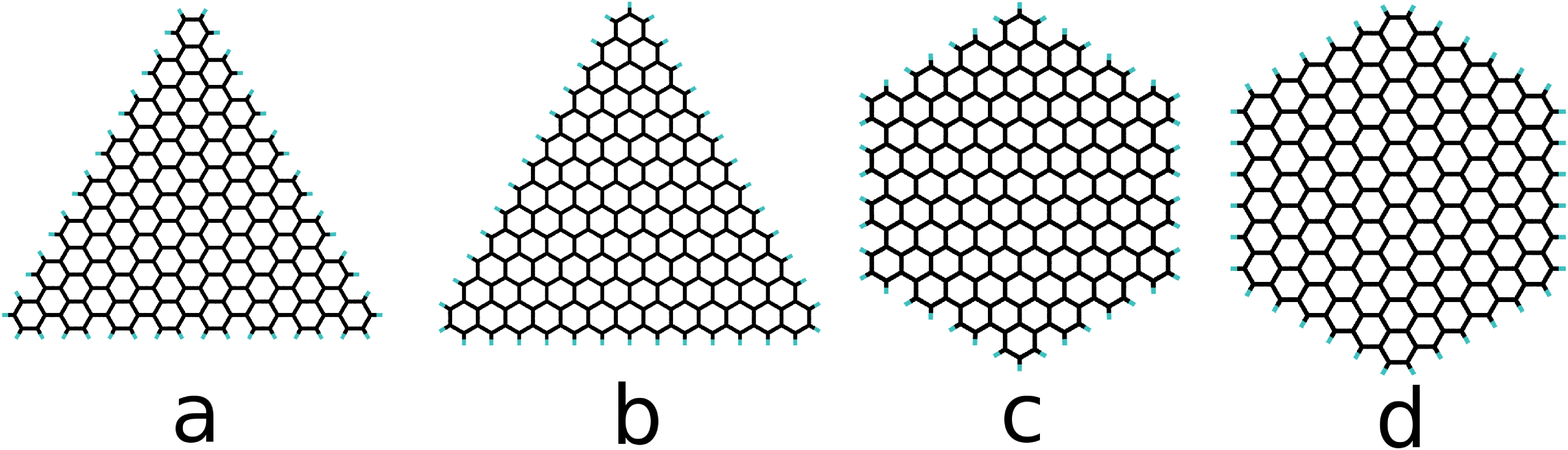}}
\caption{\label{fig:espectros} Optical absorption spectra calculated for (a) triangular armchair; (b) triangular zigzag; (c) hexagonal armchair; (d) hexagonal zigzag GNF. Results are shown for structures of varying size, the number of atoms of each structure is detailed in the plot as N.}
\end{figure}

\section{\label{sec:intro}Introduction}

Graphene, a 2D sheet of carbon atoms arranged in a honeycomb lattice, has arguably been the most promising and investigated material since it was first isolated in 2004 \cite{Novoselov2004}. Due to its remarkable properties such as unusually high current capacity \cite{Bolotin2008}, mechanical strength \cite{Lee2008} and thermal conductivity \cite{Balandin2008}, several applications have been conceived for this material. However, its zero band gap and semi-metal character represent a limitation that needs to be sorted out for carbon-based electronics to be feasible. Some ways to open up a gap are nanostructuring graphene into graphene nanoribbons (GNR) and nanoflakes (GNF) or carbon nanotubes \cite{Ma2012,Kilina2007} as well as chemical functionalization \cite{Yan2010}. Fortunately these compounds have been extensively studied from the experimental point of view, both for GNR \cite{Li2008, Jiao2009, Kosynkin2009, Cai2010} and armchair GNF of triangular \cite{Treier2011}, hexagonal\cite{Doetz2000, Zhi2008}, and other shapes\cite{Liu2011,Nano-flakes2010}.

Quasi-zero-dimensional nanostructures derived from graphene such as GNF (also known as called graphene quantum dots) are good candidates for the aforementioned applications. The quantum confinement caused by reduced dimensionality of these structures opens a tunable bandgap \cite{Yan2010,Neubeck2010}. Several studies have characterized the electronic properties of GNF \cite{Ezawa2007,Torres,Manjavacas2013,Guclu2010,Zhang2008,Ozfidan2014,Schumacher2011,Chen2010,Shi2012b,Ma2012,Li2015,Hu2014,Heiskanen2008,Akola2008,Shi2012a,Tachikawa2011,Silva2010,Thongrattanasiri2012,Yoneda2009,Banerjee2008,Akhukov2012,Kuc2010,Fernandez-Rossier2007} and some have shed light on their optical features \cite{Manjavacas2013,Guclu2010,Zhang2008,Ozfidan2014,Schumacher2011,Li2015,Silva2010,Thongrattanasiri2012,Li2014,Wang2011,Yamijala2015,Yin2012, Cocchi2014}.
Yamijala {\em et al.} \cite{Yamijala2015} performed a systematic study of the linear and nonlinear optical properties for $\sim$400 graphene quantum dots at the ZINDO/S semiempirical level.  However, there is not, to the best of our knowledge, a study relating the optical features with structural and electronic aspects, for various shapes, sizes and types of edges.

Actually, much of what is known for GNR is related to the properties of GNF: in armchair GNR there is always a band gap, which can be attributed to edge effects,
while for zigzag nanoribbons there are always zero-energy states mainly confined along the edges, causing a peak of the density of states at the Fermi level \cite{Torres}. These edge states have been experimentally observed \cite{Kobayashi2005}. 
Some of these features also hold for GNF: armchair flakes show a band gap,although in zigzag edges gap states occur only in triangular flakes \cite{Ezawa2007,Hu2014}. Moreover, the electronic \cite{Son2006a,Palma2015}, magnetic \cite{Hod2008} and optical properties \cite{Cocchi2012, Cocchi2012a,Prezzi2008,Yang2007} of GNR are well understood from the theoretical point of view.

There seems not to be a general agreement on whether edge and corner terminations affect the electronic properties of GNF. On the one hand, it was shown that the edges and corners in GNF have little influence on the distribution of the highest occupied molecular orbital (HOMO) \cite{Shi2012b}, although the Fermi level of hexagonal flakes is independent on the type of edges but rather sensitive to corner reconstructions \cite{Barnard2011}. On the other hand, zigzag edges lower the band gap with respect to armchair edges in GNF \cite{Kuc2010} and the electronic density of states is modified by edge and corner geometries even in GNF with the same shape and similar size, both for hexagonal \cite{Barnard2011} and triangular structures \cite{Shi2012a}. 

In this work we show calculated optical absorption spectra of triangular and hexagonal GNF for a range of sizes and edge structures, and give an explanation based on the electronic structure for the different trends. Both edge and corner terminations effects on the excitation energy spectrum are analyzed. Results suggest that previously characterized edge states are responsible for the differences observed in the calculated absorption spectra.

\section{\label{sec:methods}Computational Method}

In this work we have studied four classes of hydrogen passivated GNF with different shape, edge and size, namely hexagonal and triangular flakes with armchair and zig zag edge. A representative structure for each of these systems is shown at the bottom of Figure \ref{fig:espectros}. Throughout the paper we have defined $N$ as the total number of atoms, and adopted the following abbreviation for the structures: TAC for triangular armchair, TZZ for triangular zigzag, HAC for hexagonal armchair and HZZ for hexagonal zigzag.

The electronic structure of the system was described using the self-consistent charge density functional tight binding (SCC-DFTB) method, which is based on a second-order expansion of the Kohn-Sham energy around a reference density of the neutral atomic species \cite{Frauenheim2002}. This method has been extensively used for the study of graphene and graphene-based structures \cite{Kuc2010,Barnard2011,Shi2012a,Shi2012b,Yamijala2015,Shi2013} and has been benchmarked with respect to DFT for several graphene based systems with defects, with bond length discrepancies around 2\% and formation energies matching DFT values within 1.5\% \cite{Zobelli2012}. We have used the DFTB+ implementation of DFTB \cite{Aradi2007}, with the \textit{mio}-1-1 set of parameters \cite{Elstner1998}, to obtain the hamiltonian, overlap matrix and the initial ground state (GS) single-electron density matrix. All purely electronic properties such us the (total and projected) density of states and HOMO-LUMO energy gaps are calculated from these results.

The use of the DFTB hamiltonian allows for the calculation of the optical properties of very large systems (results for up to 1400 atoms are shown in this work) that would be unfeasible using ab-initio methods. The DFTB model provides a well tested approximation to the self consistent electronic structure and its response to external electric fields which can however deal with systems containing thousands of atoms.

It was been shown that structures with zigzag edge are magnetic\cite{Wang2008,Hod2008,Son2006,Palma2015}. Therefore, spin-polarized calculations were carried out for all the TZZ systems and the two largest HZZ systems using the DFTB implementation for collinear spin polarization\cite{Frauenheim2002}. The rest of the them present no net polarization in the GS. The total spin polarization obtained agree to those for which there exists data in the literature\cite{Wang2008}. Structural optimization was carried out by minimizing the force restricting the $z$ coordinate to keep a planar configuration.

The methodology applied for the calculation of optical properties is based on the real-time propagation of the density matrix of the system, and has been successfully used in our group to study a range of molecular and nanostructured systems \cite{Primo2014,Negre2013a,Fuertes2013,Oviedo:2012ct,Negre:2012fm,Negre:2010im,Oviedo:2010ek,Negre:2008fs,Oviedo:2011jt}. A Dirac-delta-pulse-shaped perturbation ($\hat{H}=\hat{H}^0+E_0\delta(t-t_0)\hat{\mu}$) is applied to the initial GS density matrix, which then evolves according to the Liouville-von Neumann equation of motion in the non-orthogonal basis \eqref{eq:lvn} which is numerically integrated. Here $\hat{H}$ denotes the hamiltonian matrix, $S$ the overlap matrix and $\hat{\rho}$ the density matrix.
\begin{eqnarray}
\label{eq:lvn}
\frac{\partial \hat{\rho}}{\partial t} = \frac{1}{\mathrm{i}\hbar}(S^{-1}\hat{H}[\rho]\rho-\rho\hat{H}[\rho]S^{-1})
\end{eqnarray}
For spin polarized systems both up and down spin density matrices are propagated within the description provided by the DFTB model. The time step for the integration was set to $0.005 fs$. 20000 steps of dynamics were done, simulating a total time of 100 fs. The intensity of the electric field was $E_0=0.001$ V \AA$^{-1}$. In the linear response regime the dipole moment can be calculated as shown in equation \eqref{eq:dip}, where $E(\tau)=E_0 \delta(\tau-t_0)$ and $\alpha(t-\tau)$ is the polarizability along the axis over which $E(t)$ is applied.
\begin{eqnarray}
\label{eq:dip}
\mu(t)=\int_{- \infty}^\infty \alpha(t-\tau)E(\tau)\mathrm{d}\tau
\end{eqnarray}
The absorption cross-section is proportional to the imaginary part of the frequency dependent polarizability, which is obtained from the Fourier transform of eq. \eqref{eq:dip}, leading to eq. \eqref{eq:alpha} We have damped the signal with a damping factor of 0.1 fs$^{-1}$ to obtain uniform broadening of the peaks. The absorption spectra are calculated averaging the polarizabilities over the three cartesian directions.
\begin{eqnarray}
\label{eq:alpha}
\alpha(E)=\frac{\mu(E)}{E_0}
\end{eqnarray}
Triplet excitations where obtained by the same method but applying a magnetic instead of electric initial pulse. The spin-unpolarized, spin-polarized singlet and spin-polarized triplet spectra are shown in Figure S5 for the TZZ flake of 264 atoms to show the importance of including the spin polarization\cite{suppdata}. Throughout the rest of this work, spectra of spin-polarized systems only consider singlet transitions, since triplet transitions have negligible intensity.

\begin{figure}[ht]
\subfigure[]{\includegraphics[width=40mm]{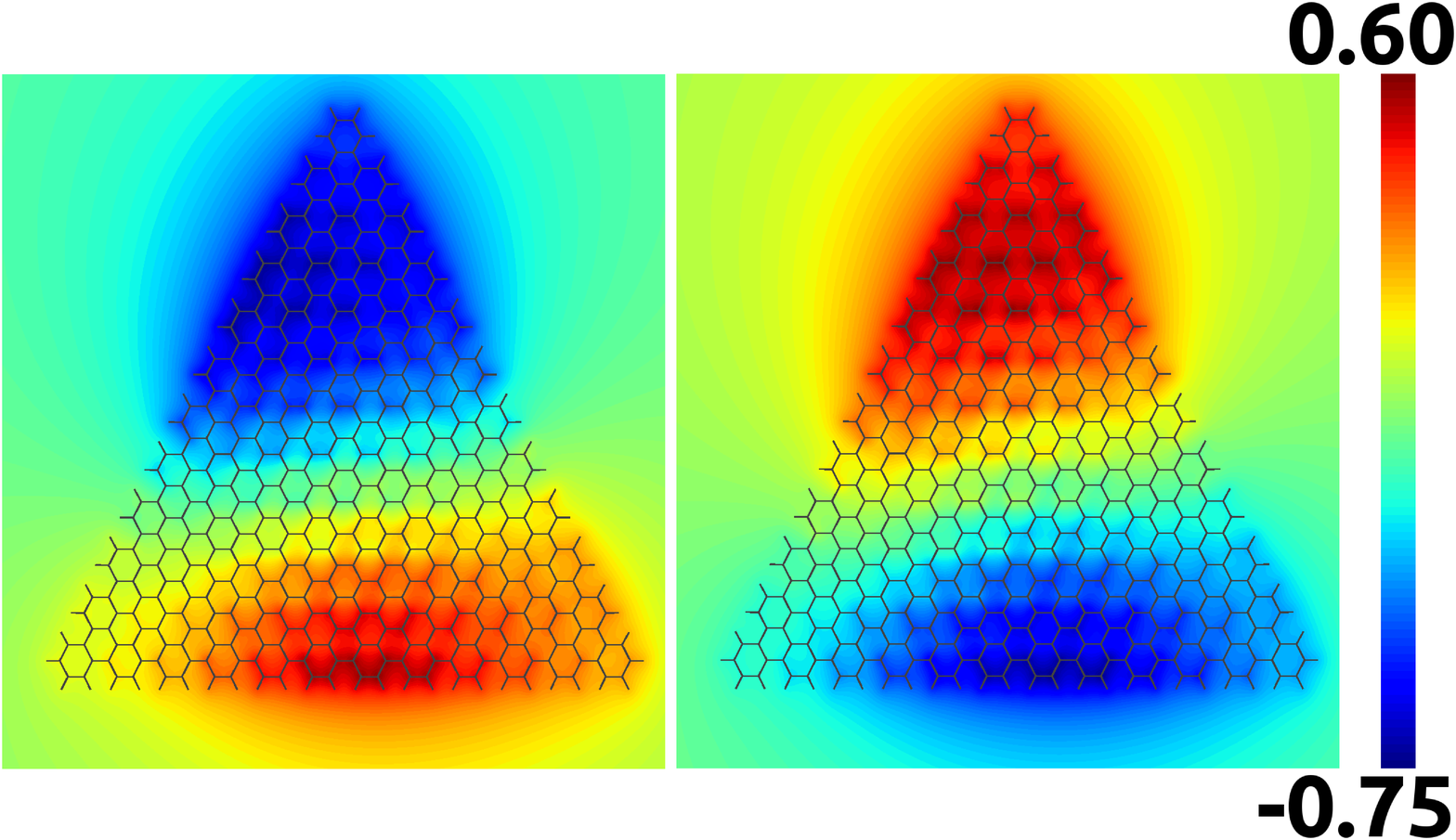}}
\subfigure[]{\includegraphics[width=40mm]{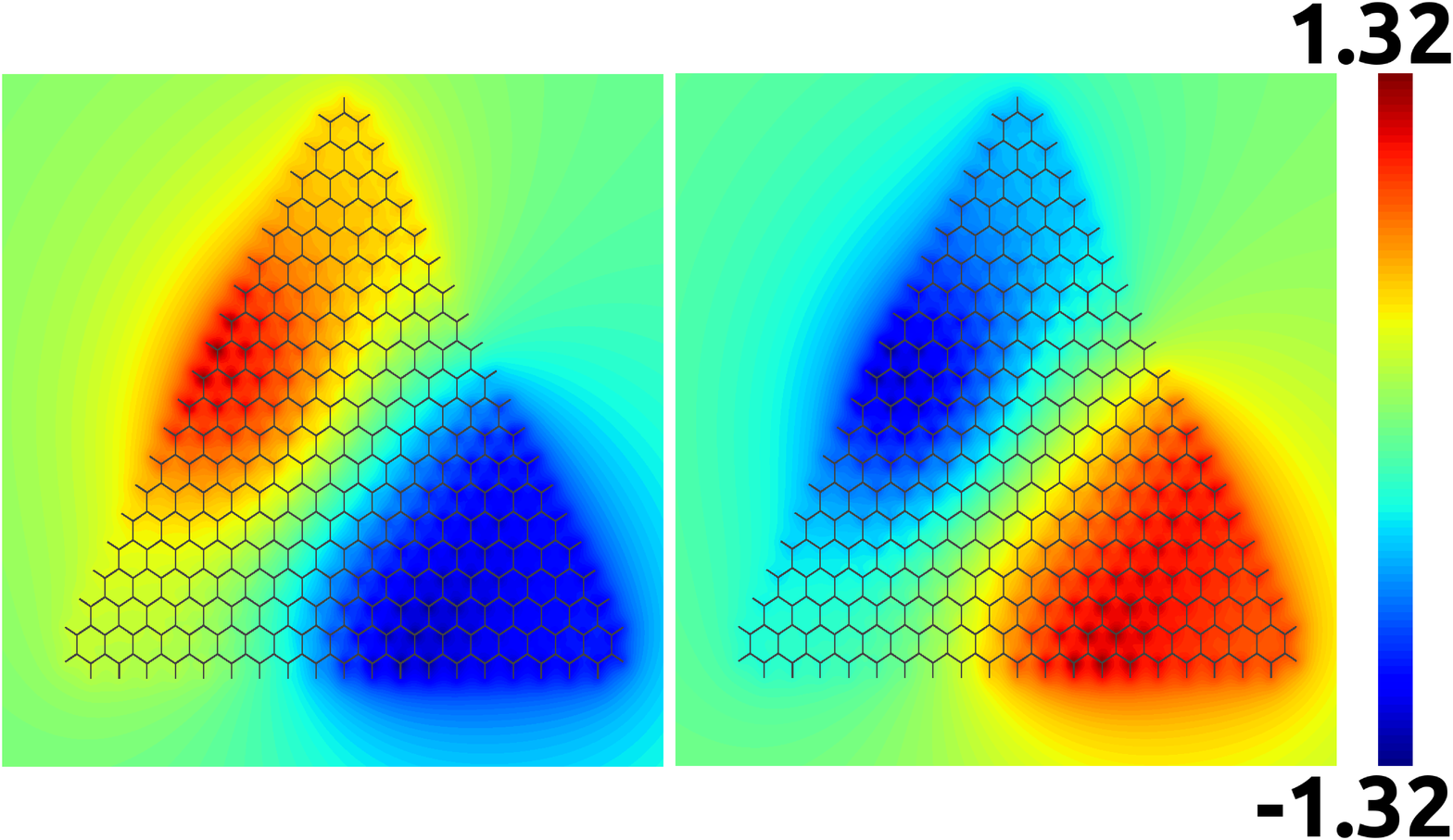}}
\subfigure[]{\includegraphics[width=40mm]{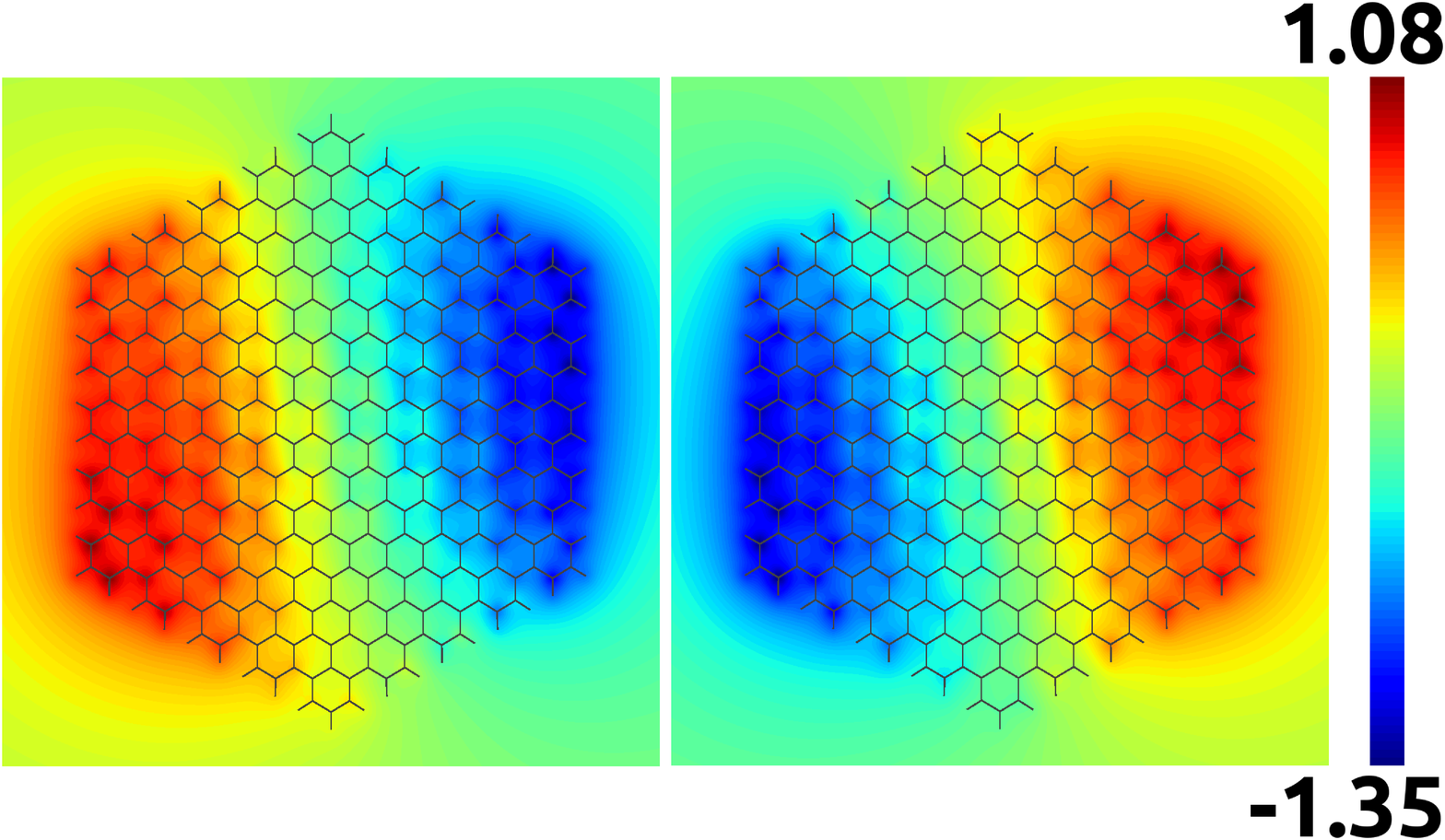}}
\subfigure[]{\includegraphics[width=40mm]{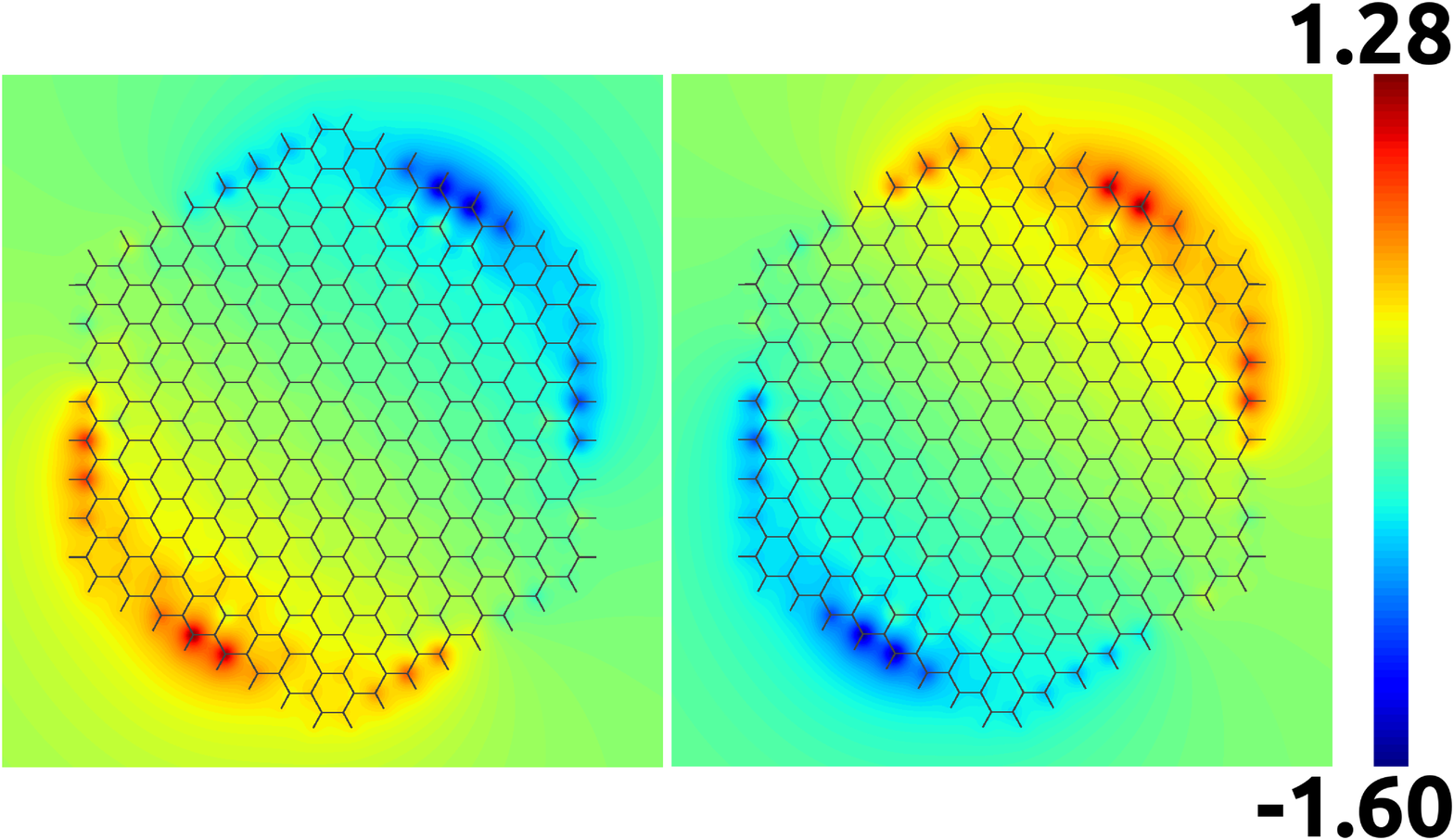}}
\caption{\label{fig:potential} Electrostatic potential at maximum and minimum dipole moment, calculated for a 50 fs propagation of the system, using a sinusoidal electric field resonant with the first peak of the spectra. The field was polarized along the maximum polarizability direction. Each of the plots correspond to (a) TAC, (b) TZZ, (c) HAC and (d) HZZ GNF with 540, 498, 420 and 432 atoms, respectively.}
\end{figure}

\section{\label{sec:results}Results and Discussion}

 \subsection{\label{sec:optical}Optical absorption spectra}
 
The optical absorption spectra of zigzag and armchair, triangular and hexagonal GNF, with sizes ranging from 2.9 nm to 7.3 nm, are shown in Figure \ref{fig:espectros}. It has been reported that the lowest-energy excitation usually involves delocalized $\pi$ electrons \cite{Torres}. Supplemental Figure S1 shows a comparison of the total density of states (DOS) and projected DOS on $p_z$ orbitals for the states involves in these transitions, supporting the $\pi$ character\cite{suppdata}. Further evidence of collective excitation can be obtained from the electrostatic potential that arises when a resonant field is applied resonant to the lowest-energy signal, plotted for the maximum and minimum dipole moment in Figure \ref{fig:potential}, for one structure of each of the four types. Section S2 (Supporting Information) contains the details of how this calculation is done\cite{suppdata}. As it is evident, charge polarization is spread in the surface for HAC, TAC and TZZ structures, while it is only delocalized in the edges for the HZZ flake.

\begin{figure}[ht]
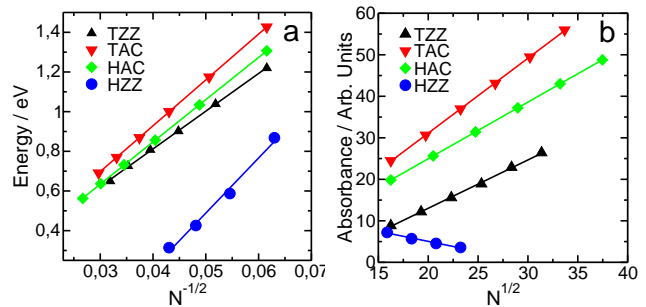

\subfigure{\includegraphics[width=42mm]{frec-vs-n12-all}}
\subfigure{\includegraphics[width=40mm]{int-vs-n12-all}}
\caption{\label{fig:frecint} First excitation energy as a function of the inverse square root of the number of atoms (a). Intensity as a function of the square root of the number of atoms (b).}
\end{figure}

A comparison of the main spectral features with results reported elsewhere can be enlightening. Cocchi {\em et al.} also calculated absorption spectra for GNF of the HAC family, albeit for smaller structures, using the ZINDO/S model based on configuration interaction including single excitations\cite{Cocchi2014}. A result for the flake of 264 atoms can be directly compared with the spectrum in Figure \ref{fig:espectros} (c). They observe the first optically active transition at 2.38 eV and report two dark transitions that are nevertheless detected in experiments. We observe transitions at energies close to where these dark transitions should appear.

A redshift of the lowest-energy excitation occurs as the GNF size increases, since the HOMO and LUMO approach as the electronic structure tends to that of bulk graphene. However, different trends in the peak intensity are observed: while for triangular (Fig. \ref{fig:espectros} a and b) and armchair hexagonal (Fig. \ref{fig:espectros} c) flakes the intensity increases, for zigzag hexagonal structures (Fig. \ref{fig:espectros} d) it decreases. The first and second peaks for HZZ structures can be told apart given the fact that they follow different trends, as it can be drawn from Figure S2\cite{suppdata}. The intensity of the first peak for the two largest structures is so weak that it could not be detected, and hence could not be included in the forthcoming analysis.

Figure \ref{fig:frecint} offers further insight on the observed trends. First of all, the relation between the excitation energy and the inverse square root of the number of atoms $N$ is linear, which resembles the same type of relation for collective excitations in metallic nanoparticles with $N^{-1/3}$, namely, the surface/volume ratio\cite{Negre:2008fs}. Hence, for GNF the lowest excitation energies depend linearly on the perimeter/surface ratio, a typical quantum confinement feature. Secondly, a similar relation holds for the intensity as a function of the square root of the number of atoms, as a consequence of the increasing polarizability linked to a larger number of electrons involved in the excitation for large structures. However, this is not the case of HZZ flakes. The reason is described in the following paragraph.

\begin{figure}[ht]
\subfigure[]{\includegraphics[width=40mm]{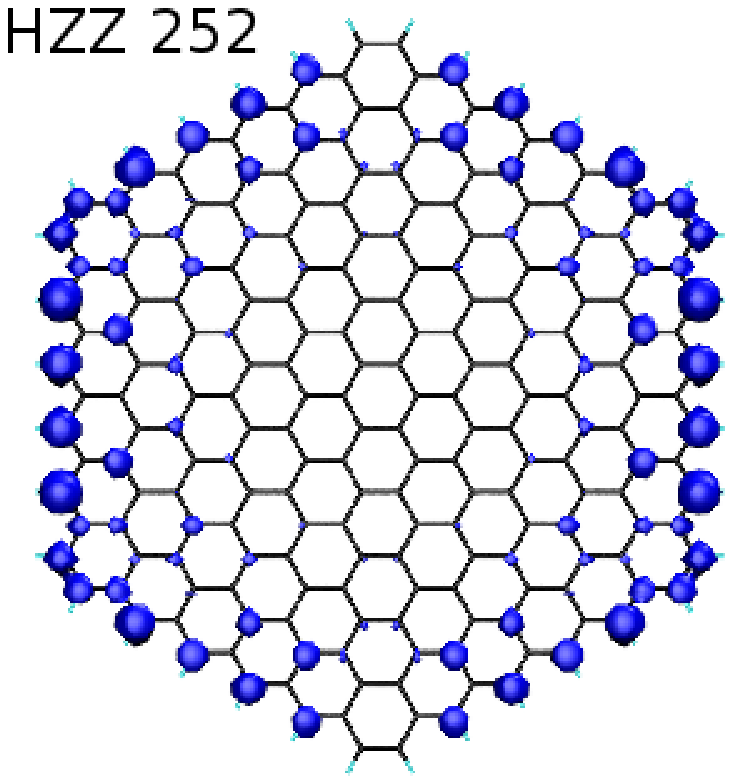}}
\subfigure[]{\includegraphics[width=40mm]{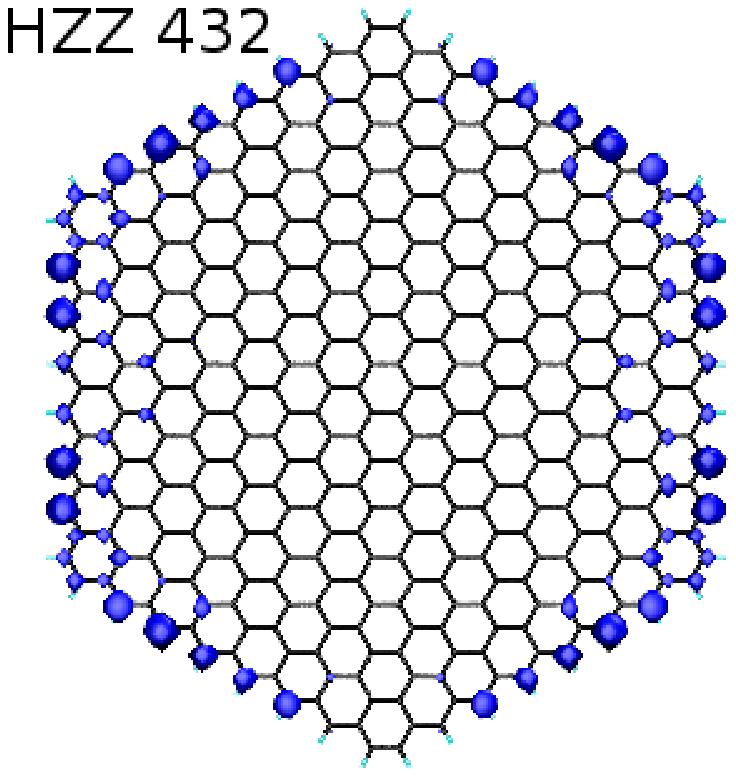}}
\caption{\label{fig:tddshzz} Transition dipole density for the first peak of the spectra for HZZ structures of (a) 252 and (b) 432 atoms, calculated as described in the text.}
\end{figure}

The edge character of the excitation observed in Figure \ref{fig:potential} for HZZ flakes suggests that only edge states participate in the transition; by inspecting the projected density of states per atom (see section \ref{sec:electronic} of a description), the number of such states was found to be 12 one-particle states, half of them filled at 0K, for all the structures. A visual representation of such states can be found in Figure S3 \cite{suppdata}. As a consequence, as the GNF size increases the dipole density decreases, since the a constant number of electrons is spread in an increasing edge perimeter. Visual evidence of this phenomenon is obtained by plotting the transition dipole density (TDD). To calculate the TDD, we consider only the two orbitals that contribute the most on this transition, $\phi_{initial}$ and $\phi_{final}$, which are obtained as described in Section S5 and Figure S4\cite{suppdata}. A real space picture of the TDD is calculated as 
\begin{equation}
\begin{split}
\rho(\mathbf{r}) = &|\phi^*_{final}(\mathbf{r}) \hat{x} \phi_{initial}(\mathbf{r})|^2+ \\
& |\phi^*_{final}(\mathbf{r}) \hat{y} \phi_{initial}(\mathbf{r})|^2+ \\
& |\phi^*_{final}(\mathbf{r}) \hat{z} \phi_{initial}(\mathbf{r})|^2
\end{split}
\end{equation}
where $\mathbf{r}=(x,y,z)$. It is essentially a spatially weighted inner product between initial and final orbitals. In Figure \ref{fig:tddshzz}, a comparison of $\rho$ for two HZZ structures is done, where the same density isosurface value has been considered. The decrease in the transition dipole moment for the larger structure is evident, which in turn explains the decreasing absorption cross section. 

\subsection{\label{sec:electronic}Electronic structure }

\begin{figure}[ht]
\subfigure[]{\includegraphics[width=40mm]{dos-dirac-tac1134}}
\subfigure[]{\includegraphics[width=42mm]{dos-dirac-tzz984}}
\subfigure[]{\includegraphics[width=40mm]{DOS-dirac-hac1104}}
\subfigure[]{\includegraphics[width=42mm]{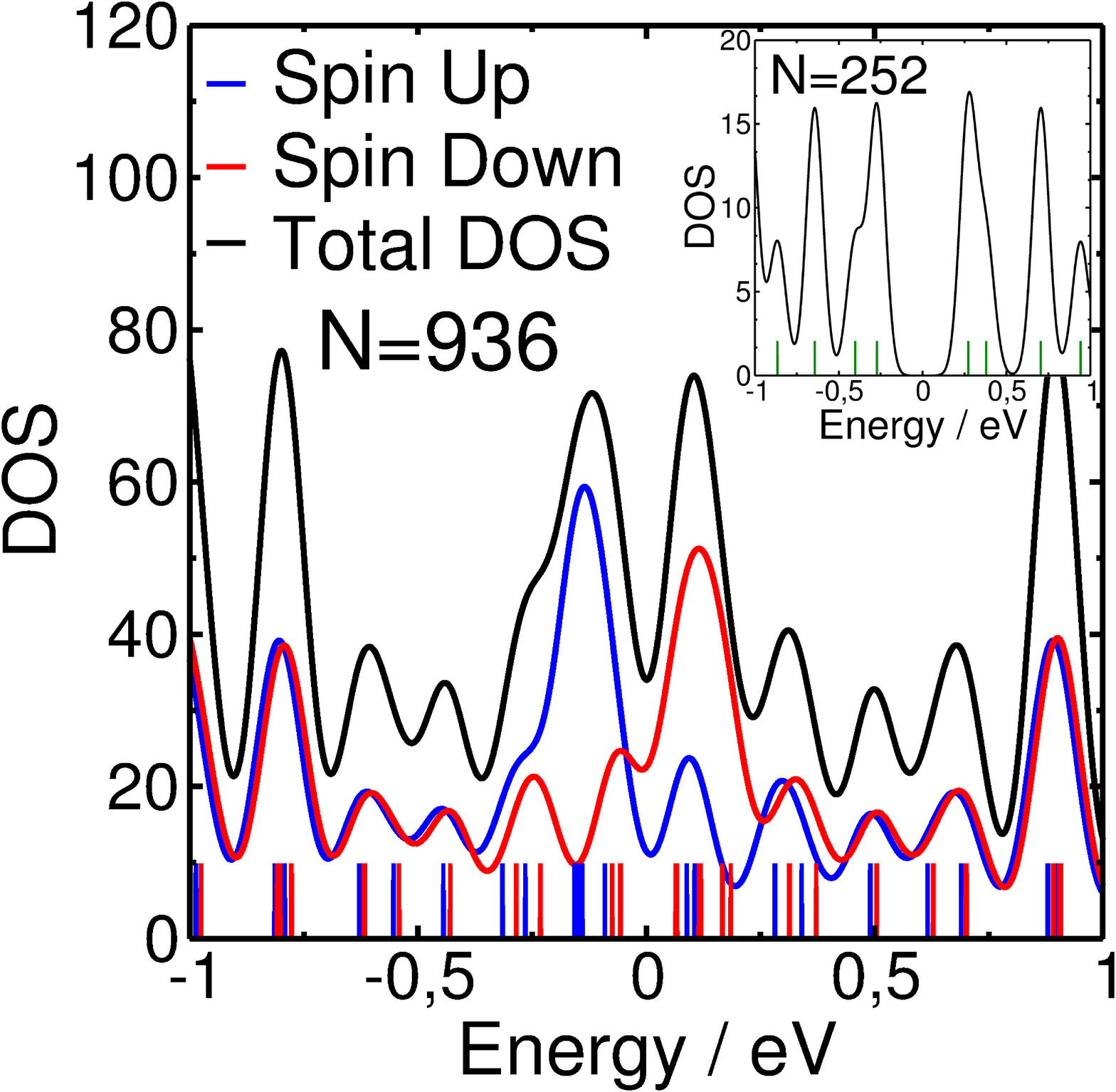}}
\caption{\label{fig:dos} Density of states (black curve) and discrete energy spectrum (green, blue and red bars) of a (a) $N=1134$ triangular armchair; (b) $N=984$ triangular zigzag; (c) $N=1104$ hexagonal armchair; (d) $N=936$ hexagonal zigzag flake. The inset in (d) shows the density of states of a smaller $N=252$ hexagonal GNF, where a band gap is clearly seen.}
\end{figure}

The main features that arise from time-dependent calculations are consistent with what is expected from electronic structure analysis. The density of states (centered at the Fermi level $E_F$) shown in Figure \ref{fig:dos} was plotted with a Gaussian envelope at each eigenenergy $\epsilon_i$ according to 
\begin{eqnarray}
DOS(\epsilon)=\frac{1}{\sqrt{2 \pi} \Gamma}  \sum_i \exp(-(\epsilon-\epsilon_i)^2/2\Gamma^2).
\end{eqnarray}
where $\epsilon$ denotes the energy and $\Gamma$ is the standard deviation of the distribution (a value of $\Gamma=0.05$ eV was used in all cases).
A HOMO-LUMO gap around $E_F$ is found in armchair, caused by the same reason than in armchair GNR \cite{Torres}. Zigzag triangular flakes show several ``gap states'' close to $E_F$, which in spin-unpolarized calculations appear as zero-energy states whose multiplicity increases with size\cite{Ezawa2007}. However it has been proven that these nonbonding states arise due to topological frustration (impossibility for all $\pi$ bonds to be satisfied simultaneously). When the spin polarization is included, the asymmetry of spin breaks the degeneracy opening a small gap at the Fermi level \cite{Ezawa2007,Wang2008}. Furthermore, the total spin ground state can be very large since it scales linearly with size \cite{Wang2008}, as it is shown in Figure S6\cite{suppdata}.

Zigzag hexagonal flakes, on the other hand, show a continuous distribution of states with no clear energy gap. In the latter, the density of states close to the Fermi level increases as the flake becomes larger (Figure \ref{fig:dos} d), giving rise to very distinct DOS profiles: a small gap is present for smaller $N=252$ flakes (inset), but a continuous DOS is seen around $E_F$ for larger ones ($N=936$). This phenomenon is visible for structures larger than $\approx$ 500 atoms, which are also magnetic. Transitions among these states are responsible for low-energy, low-intensity transitions in the absorption spectra.

In general, the calculated DOS are consistent with previous calculations using a simple nearest-neighbor, one-orbital Hückel model \cite{Heiskanen2008,Akola2008,Zhang2008,Fernandez-Rossier2007} and DFT \cite{Hu2014,Silva2010}. For structures comparable to the ones analyzed here, usually a wide range of energies is considered in the literature, which have very similar DOS profiles among the different structures \cite{Barnard2011}. On the contrary, we observe a remarkable difference in the lowest electronic states for different type of edges. This implies that a close inspection of this range of energies is crucial for a correct understanding of the practically relevant optical features.

\begin{figure}[ht]
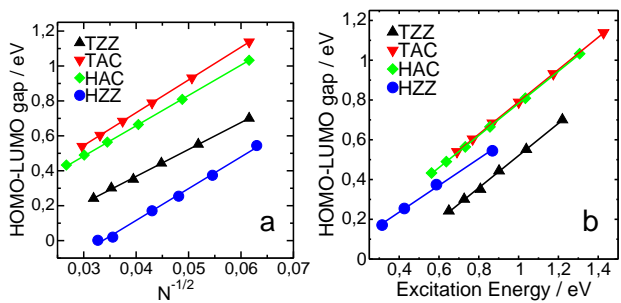

\subfigure{\includegraphics[width=40mm]{homolumo-vs-N12}}
\subfigure{\includegraphics[width=40mm]{homolumo-vs-E}}
\caption{\label{fig:gaps}HOMO-LUMO gaps as a function of (a) the inverse square root of the number of atoms and as inverse of number of atoms (inset) only for HZZ.(b) HOMO-LUMO gaps as a function of the excitation energy.}
\end{figure}

The trends shown in Figure \ref{fig:frecint} are directly related to the dependence of the HOMO-LUMO gap with structural properties. In Figure \ref{fig:gaps} (a) the correlation between the gap energy and the inverse square root of the number of atoms is depicted. The general observation is a decrease in the gap energy with increasing the size due to quantum confinement \cite{Berger2006}, also found in previous calculations \cite{Zarenia2011,Yamijala2015,Zhang2008}. This explains the red-shift in the absorption spectra. The higher excitation energies values seen in armchair with respect to zigzag flakes also holds when HOMO-LUMO gaps are considered instead of excitations (Figure \ref{fig:gaps}), and agrees with previous calculations \cite{Kuc2010}. There is a linear relation between the HOMO-LUMO gaps and the excitation energies (see Figure \ref{fig:gaps} (b)), showing that the self-interaction term of the excitation density is small, due to the spatial delocalization of these excitations\cite{MBOviedo}.
A natural classification in three groups arises by observing this plot, suggesting that although armchair flakes have very similar electronic properties, zigzag flakes have intrinsic shape-dependent features.

\begin{figure}[ht]
\includegraphics[width=0.45\textwidth]{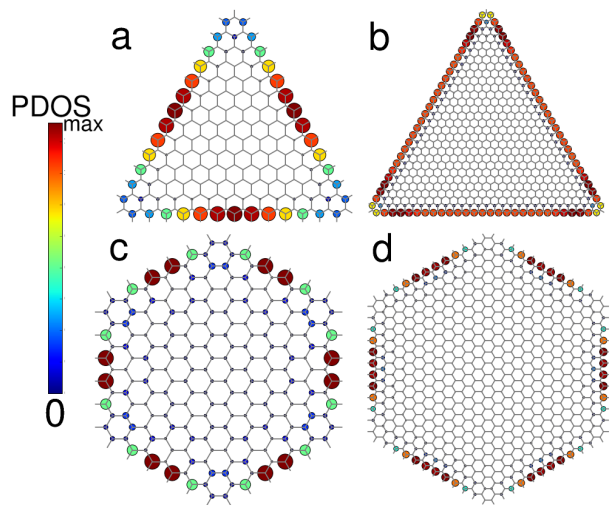}
\caption{\label{fig:pdos} Normalized projected density of states (PDOS) per atom at $E_F$ of (a) $N=264$ and (b) $N=984$ triangular zigzag; (c) $N=252$ and (d) $N=936$ hexagonal zigzag GNF. The area of the circle, as well as the color, is proportional to the contribution of each atom to the PDOS at the $E_F$. $PDOS_{max}$ for each structure is (a) 0.012 eV$^{-1}$; (b) 1.2 eV$^{-1}$; (c) 0.27 eV$^{-1}$; (d) 0.54 eV$^{-1}$}
\end{figure}

To explore the nature of low-energy states in zigzag GNF the projected density of states (PDOS) contribution per atom was calculated. The PDOS of orbital $\alpha$ in a non-orthogonal basis set is calculated according to
\begin{eqnarray}
PDOS_\alpha(\epsilon) = \sum_i \frac{|(SC)_{i \alpha}|^2}{\sqrt{2 \pi} \Gamma} \exp(-(\epsilon-\epsilon_i)^2/2\Gamma^2)
\end{eqnarray}
where $S$ and $C$ are the overlap and the molecular orbital coefficient matrices respectively. The PDOS is analogous to the DOS but weighted by the factor $|(SC)_{i \alpha}|^2$ indicating the contribution of localized orbital $\alpha$ to molecular orbital $i$. We observe that carbon $p_z$ orbitals are the only ones that contribute to the total DOS close to the Fermi level\cite{suppdata} (see Figure S1), which explains the success of one-orbital tight-binding or Hückel models for the study of these systems \cite{Heiskanen2008,Akola2008,Zhang2008}.
Figure \ref{fig:pdos} shows the sum of the contributions of all orbitals of each atom to the state at (or close to) $E_F$ for zigzag triangular (a-b) and hexagonal (c-d) GNF. It is evident that such states are mainly localized on the edge atoms, and are hence edge states. For TZZ flakes, the lowest-energy absorption peak involves transitions from edge states to delocalized states, which give rise to a delocalized electrostatic potential profile as shown in Figure \ref{fig:potential} (b). For HZZ structures, such transition involves only edge orbitals, as evidenced in Figure \ref{fig:potential} (d).

Moreover, there is a strong difference in the magnitude of the contribution of each atom to the total DOS between zigzag triangular and hexagonal GNF. In the former, the PDOS per atom increases by two orders of magnitude from $N=264$ to $N=936$, while in latter the PDOS only doubles from $N=252$ to $N=936$. This difference in the hexagonal structures gives rise to the different profiles in the DOS plots shown in Figure \ref{fig:dos} d. The origin of these states has been identified in GNR \cite{Nakada1996} and explained for GNF \cite{Ezawa2007,Zhang2008}. Due to the gap around $E_F$ in armchair GNF, this analysis cannot be done for such structures. It is clear that the frontier orbitals have large edge dependence in zigzag flakes. This is key for the understanding of the optical properties of such structures.

\subsection{\label{corners}Effect of corner geometry}

\begin{figure}
\subfigure{\includegraphics[width=41mm]{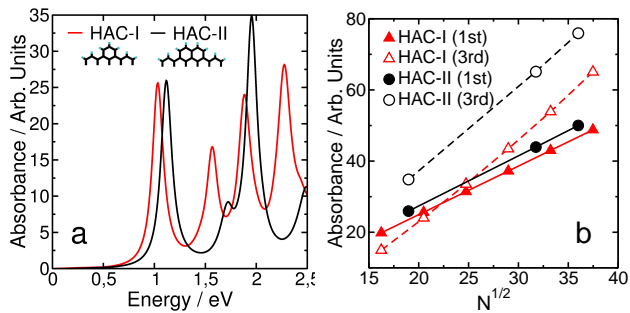}}
\subfigure{\includegraphics[width=40mm]{todos-1er3er-corr}}
\caption{\label{fig:spechex} (a) Absorption spectra for armchair hexagonal flakes with type I ($N=420$, red line) and type II corners ($N=360$, black line). (b) Intensity as a function of square root of number of atoms for the first (continuous line) and third (dashed lines) excitations in armchair hexagonal flakes with type I (red) and type II corners (black).}
\end{figure}

\begin{figure}
\subfigure{\includegraphics[width=40mm]{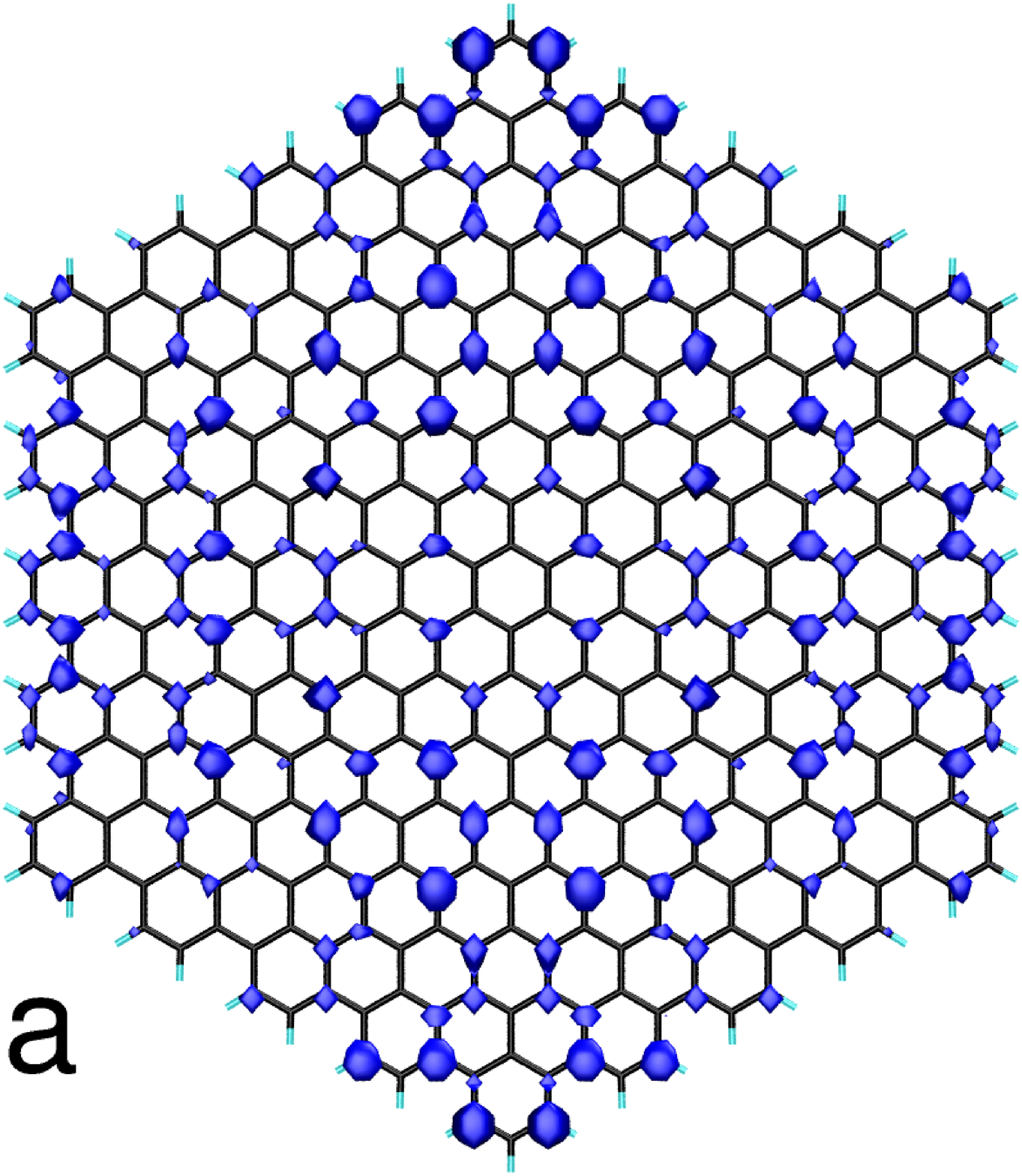}}
\subfigure{\includegraphics[width=40mm]{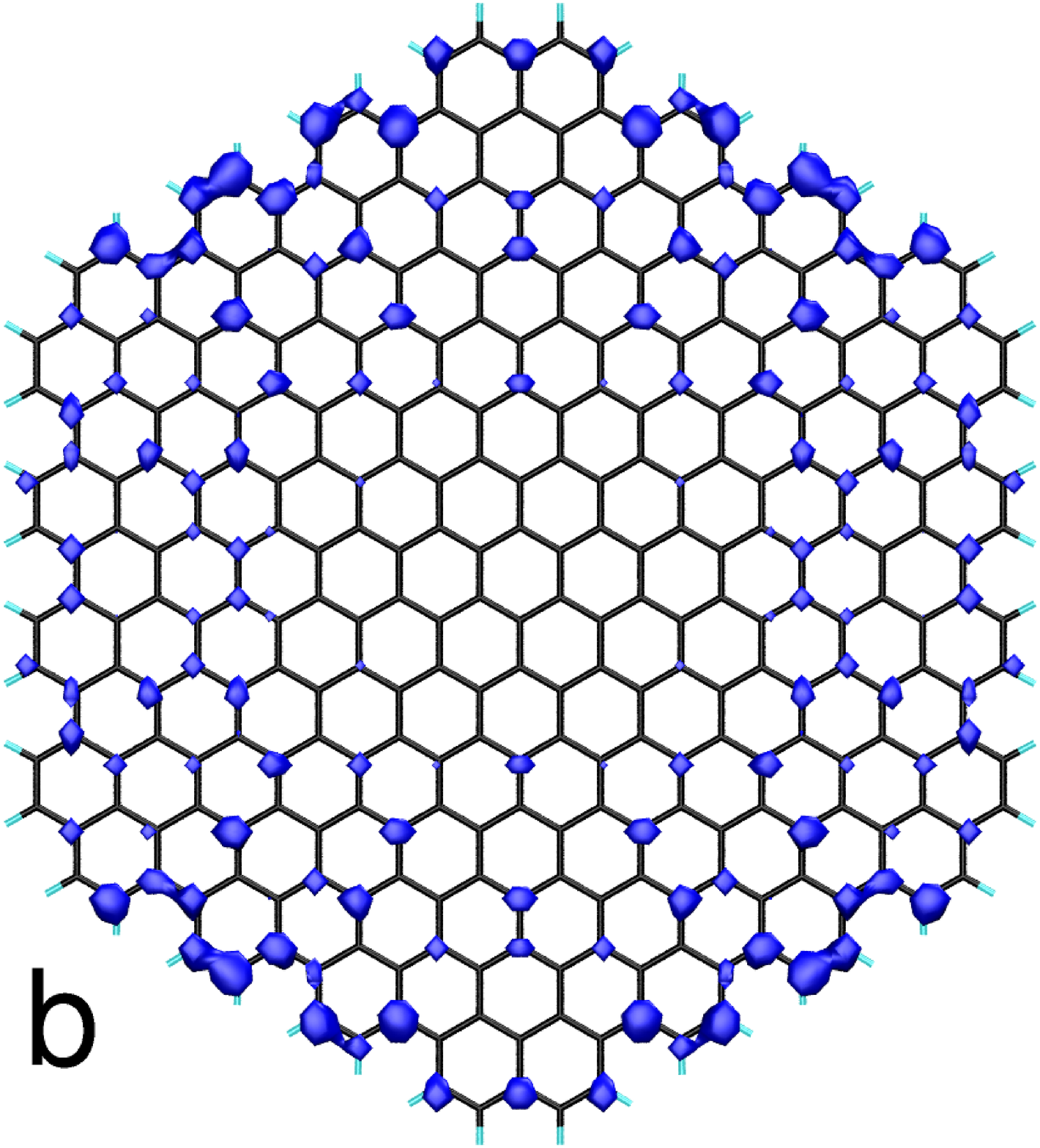}}
\caption{\label{fig:doshex} Transition dipole density of the transition corresponding to the third peak of the absorption spectra for armchair hexagonal (a) type I ($N=420$) and (b) type II ($N=360$) corners.}
\end{figure}

Barnard and Snook \cite{Barnard2011} examine the possible reconstructions that corner atoms undergo in two corner terminations of armchair hexagonal GNF, analyzing the forces and the density of states in a wide range of energies. However, low-energy electronic aspects needs to be considered in order to explain optical properties.
In Figure \ref{fig:spechex} (a) the spectra of two armchair hexagonal nanoflakes (similar in size but not equal) with two possible (zigzag) corner terminations, ending in one benzene ring (HAC-I) or two benzene rings (HAC-II) are shown. The spectra are similar up to the lowest-energy peak, but show different profiles for higher energies. The comparison between the size dependence of the excitation intensity of the first and third peaks is done in Figure \ref{fig:spechex} (b). Although they show similar intensity 
for the first peak, when the third peak is considered the intensities differ considerably. A larger absorption intensity is caused by a large transition dipole moment. To gain insight on the structural dependence of this stronger absorption, we calculated the transition dipole density for these strcutures. The same isovalue of $\rho(x,y,z)$ for the orbitals involved in the third excitation is shown in Figure \ref{fig:doshex} for both structures analyzed previously. It is evident from the larger volume enclosed by the isosurfaces, that the transition density is highly polarized in the HAC-II structure (Figure \ref{fig:doshex} (b)), giving rise to the stronger optical absorption see in Figure \ref{fig:spechex}.

\section{\label{sec:conclusions}Conclusions}

In this work we demonstrate that shape and edge geometry do matter when optical properties of GNF are considered. In fact, GNF with different geometries could be identified by their optical absorption spectra which show very distinct profiles and excitation energies. A classification in three families (A,B,C) arises from this analysis: armchair GNF belong to group A, zigzag triangular GNF to group B and zigzag hexagonal GNF to group C. Group A GNF show first excitation peaks with linearly increasing intensity and decreasing energy with respect to size, and higher excitation energies are expected since the HOMO-LUMO gap is larger. In group B the trends in intensity and energy are the same than group A, but the excitation energies are smaller because of gap states. In group C intensity decreases with size because of the edge dependence and constant number of involved states. Different corner terminations do not affect this classification since they become relevant when higher excitation energies of the spectrum are analyzed.

Given the structural polydispersity that is naturally achieved in prepared experimental samples of these nanomaterials at present days, studies that contribute to the understanding of relevant properties in different structural configurations are imperative. The design of either robust materials against disperse structural features, or samples with uniform properties or small property dispersity, are benefited from studies like this one. We show that, by performing a statistic analysis over the optical properties of a mixture of GNF structures, it could be possible to measure the proportions of GNFs with a certain shape and edge type, and tell apart two corner terminations. Further studies should be performed, nevertheless, to extend the characterization and provide useful insight for the design of novel applications.

\section{\label{sec:acknowledgments}Acknowledgments}

The authors acknowledge financial support by Consejo Nacional de Investigaciones Científicas y Técnicas (CONICET) through grant PIP 112-201101-0092 and wish to thank SeCyT UNC for the funding received. This work used computational resources from CCAD – Universidad Nacional de Córdoba (http://ccad.unc.edu.ar/), in particular the Mendieta Cluster, which is part of SNCAD – MinCyT, República Argentina. CMW and FPB are grateful for studentships from CONICET.

\bibliographystyle{apsrev}
\bibliography{graphene_paper}

\end{document}